# A New Multi-Tracer Approach to Defining the Spiral arm width in the Milky Way

Jacques P Vallée

National Research Council Canada, Herzberg Astronomy & Astrophysics, 5071 West Saanich road, Victoria, BC, Canada V9E 2E7,  jacques.p.vallee@gmail.com

**Abstract.**

We analyze recent observations of the spiral arm width in the Milky Way, as a function of the galactic radius, and we compare this relation with the prediction from the density wave theory.  We use the following method:  in each spiral arm, we concentrate on the separation (or offset) between the starforming region (radio masers) near the 'shock front' of a density wave, and the aged star region (diffuse CO gas) near the 'potential minimum' of a density wave; we take this separation between these two tracers as the arm width.  New results: we find a typical separation (maser to diffuse CO gas) near 250 ± 50 pc, and an increase of this separation with galactic radius of about 25  ± 5  pc per kpc. We note that, as expected, this separation is somewhat smaller than that found earlier between the dust lane and the aged star region. Overall, these results supports the basics of a density wave.

## 1.  Introduction.

Comparisons of Milky Way structure and density wave theory are so few and far between because of the difficulties of measuring distances in our own Galaxy.  A key prediction of the density wave theory is the separation of dust and starforming regions in the inner arm edge ('shock front') from the older stars and broad diffuse gas ('potential minimum') – see Fig.2 in Roberts (1975). How can we measure that separation? Thus at least 2 arm tracers are needed.

The  prediction of a separation between the starforming region versus the region with older stars and diffuse CO  gas, within a spiral arm, in the density wave theory was made over 50 years ago (Lin & Shu 1964; Roberts 1969). The early lack of such observed separation has been suggested as an impediment: "the earliest failure, seemingly, was not detecting color gradients associated with the migration of OB stars whose formation is triggered downstream  from the spiral shock front" (Shu 2016). Over the ensuing half-centennial years, such needed precision in distances was difficult to achieve observationally, and in the mean time several alternate theories were created and proposed without a clear separation between starforming regions and aged stars.

A review of theories to explain the creation and evolution of spiral arms was given in Dobbs & Baba (2014). Many different theories have been made, in the absence of precise observations to verify their predictions, and some theories have since fallen out of favor. In the last two decades, the parallax distance became achievable observationally, mostly using radio masers. This could now allow more precise distance measurements, and to check the predicted separation between the starforming regions and the location of older stars and broad diffuse CO intensity peak.

This separation or offset between the starforming region (dust and radio masers at the inner arm edge) and the location of the potential minimum (aged stars and diffuse CO gas peak intensity) was first observed in our Milky Way disk recently (Vallée 2014a; Vallée 2014b), about 5 decades after their prediction.

A previous measurement of the separation between the dust lane and the diffuse CO 1-0 gas intensity peak (over all galactic spiral arms) was found at 315 ± 26 pc (Vallée 2016a – his Table 1 and Fig. 1). This new study suggests that the radio masers are located in between the dust lane and the diffuse CO 1-0 gas peak, 65 pc away from the dust lane for a maser-to-CO offset of 250 ± 50 pc (mean of Table2 in Section 3).

Observationally, a tangential look with a radio telescope with a broad beam (near 8') at a spiral arm in Galactic Quadrants I and IV employing the diffuse large-scale CO 1-0 tracer can yield the galactic longitudes of the diffuse CO 1-0 arm tangents. These tangent values can be compared with the mean galactic longitude of the radio masers close to a spiral arm. Comparing these longitude values, an angular separation can be obtained, and transformed into a linear separation at the arm's distance from the Sun. For a review, see Vallée (2014a – his Table 4) and Vallée 2016a (his Tables 1 to 10), with a mean separation of dust to broad diffuse CO 1-0 gas near 315 ± 26 pc. In addition, a fit of the galactic longitude of the tangent to each spiral arm can be made, using logarithmic spiral arms in both Galactic Quadrants I and IV, in order to get a global spiral arm model for the diffuse CO 1-0 gas. Such a global view, using an arm tracer in both Galactic Quadrants I and IV, allows a more precise determination of the spiral arm pitch angle. Thus for the Sagittarius and Scutum arms (Vallée 2015 – his Fig.1 and equation 10) and for the Norma arm (Vallée 2017a – his Table 1), a mean pitch angle near $-13.1° ± 0.5°$ obtained.

Nearer the galactic Meridian (the line joining the Sun to the Galactic Center), the radio masers can be observed trigonometrically to yield their parallactic distances. When comparing with the predicted location of the diffuse CO in the nearest arm, a linear separation (maser to diffuse CO gas) can be obtained. This separation was measured for the Perseus arm (Vallée 2018a), and also for other arms (Vallée 2019 – his Table 7).

In Section 2, we employ the diffuse CO 1-0 spiral arm model (Vallée 2017a and 2017b) and we compile from the literature the separation in galactic longitude between the maser locations and the diffuse CO 1-0 gas locations (Table 1), as well as the separation near the Galactic Meridian of the maser locations from the diffuse CO gas peak (Table 2), and we analyze the results (Figure 1).

In Section 3, we investigate the observed increase in this separation, when increasing the Galactic radius (Fig. 2), and compare this with the theoretically predicted change (Fig.3).

In Section 4, we investigate the distance to newly discovered radio masers beyond the Galactic Center (Table 3, Fig. 4), and show their locations in the Milky Way disk (Fig. 5) and in a kinematical map (Fig. 6).

In Section 5, we discuss these new results, in relation to the predictions of the recent maser-based model of Reid at al (2019). In Section 6, we briefly summarize recent results using a multi-tracer approach to define the spiral arm width in 40 nearby spiral galaxies. We conclude in Section 7.

## 2. The offset of masers with respect to the arm seen in diffuse CO 1-0 intensity

### 2.1 Do we need 2 different arm tracers, for each arm?

If the dust and gas distributions within an arm were homogeneous, then the arm width and arm center could be measured readily, using a single arm tracer. This old method of defining an arm width with the scatter in one tracer alone is archaic. Defining a spiral arm's width using one tracer alone is potentially very misleading.

Which tracers can be employed? Here we selected radio masers, given their better distance estimates (near the location of dust lanes), as well as the diffuse CO 1-0 gas (near the 'potential minimum'). Observationally, the location of the 'older stars' was also found to be near the 'potential minimum' (see Table 2 and Figure 4 in Vallée 2017d), thus very close to where the 'diffuse CO' is found.

Radio masers are extremely young protostars often located in extremely young HII regions in dusty areas (hence not yet seen at optical wavelengths), and thus extremely close to the shock front (dust lane). Visible optical HII regions are older and have already moved away from the shock front. Most visible HII regions are located somewhere between the shock front and the potential minimum. These various locations, based on the average for several spiral arms in the Milky Way, are shown in Table 2 and Figure 2 in Vallée (2014b), Table 6 in Vallée (2016a), Table 2 in Vallée (2017d).

From the Sun, one can observe with a telescope scanning in galactic longitude the broad diffuse CO 1-0 gas, and record the galactic longitude value of each arm (the arm tangent as seen in diffuse CO).

The 'diffuse CO 1-0 gas' mentioned in the text is that cold gas observed with a single-dish telescope beam of about 8 arcmin, hence the words 'diffuse gas' (see Table 3 in Vallée 2014b). It is often missed by a radio interferometer without some small-scale baselines.

The 'clumpy CO gas' is seen elsewhere, around the visible HII regions, with a narrower single-dish telescope beamwidth, or with a radio interferometer without small scale baselines. The H-alpha visible HII regions are not at the shock front, nor at the 'potential minimum' – they are somewhere in between. In single-dish telescope maps with large angular beams, we assume that the low-density diffuse CO gas (over a long arm line of sight) will predominate over the few high-density clumpy CO gas clouds associated with HII regions (with a shorter clump line of sight). Thus one could then distinguish the larger-scale diffuse CO gas in the spiral arm from the CO in the dense bright clouds associated with the young HII regions.

The diffuse CO 1-0 gas was observed with single-dish telescopes, most with an angular resolution of 8.8' (Columbia survey), and some with 7.5' (CfA survey) and some with 8.4' (Columbia survey) - see observational details in Table 5 in Vallée (2016a).

From these galactic longitude values, a 4-arm spiral model can be fitted to these observations (Vallée, 2017a and 2017b). This CO-fitted arm model can be employed to give the kinematical map of the gas in the Milky Way disk, assuming a rotation curve and a solar distance to the Galactic Center (Vallée 2017c), and an arm pitch angle of $-13.1°$ as found observationally (Vallée 2015). For an overview, see Vallée (2017d).

## 2.2 The location of each different arm tracer, within an arm

On the observational side, a search of the literature yielded several measurements of this offset. Several angular locations were measured in different arm tracers looking tangentially from the Sun to a spiral arm in our Milky Way galaxy – see Table 5 (CO tracers) and Table 7 (maser tracers) in Vallée (2016a).

**Table 1** shows these angular offsets between these two arm tracers, for each spiral arm, and the linear offset value at the measured galactic radius and solar distance. All these offsets are measured from the broad diffuse CO 1-0 gas to the radio maser lane; the offsets are negative in Galactic Quadrant IV, but positive in Galactic Quadrant I, and thus all offsets are pointing toward the Galactic Center in each Galactic Quadrant (giving a mirror image as predicted by the density wave theory – see Vallée 2016a). Here the mean offset (col. 4, excluding the outlier in the last row) is 241 ± 39 pc where the error estimate is the standard deviation of the mean.

Figure 1 shows these orbital offsets, one along each spiral arm seen tangentially from the Sun. The tangent in galactic longitude to each arm for the masers is shown as blue bars, while that for the tangent to each arm when using the broad diffuse CO 1-0 gas tracer is shown as gray bars. These gray bars have been previously employed to fit a global spiral arm model (Vallée, 2017a and 2017b), as reproduced here (spiral curves). It can be seen that all maser lanes are inward of each arm, toward the Galactic Center. Also, when available, the location of the tangent to the Near Infrared and Mid infrared dust lane is shown as red bars – see Vallée (2017a, his Fig. 2).

In addition, several radial offsets were measured in different arm tracers along the Galactic Meridian, using masers with trigonometric distances.

**Table 2** shows these radial offsets along the Galactic Meridian ($l=0^o$ and $l=180^o$) at the measured solar distances. All these offsets are measured from the diffuse CO gas model to the maser lane; the offsets are positive in Galactic Quadrants II and III, but negative in Galactic Quadrant I and IV; thus all offsets are pointing toward the Galactic Center in each Galactic Quadrant (another mirror image across the Sun's galactic radius – see Vallée 2019). Here the mean offset (column 7) is 250 ± 50 pc.

**Figure 1** also shows these radial offsets along the Galactic Meridian for the maser lane (blue bars), as measured from the location of the diffuse CO 1-0 gas model (spiral curves).

## 3. The width of a spiral arm, increasing with the galactic radius

Here we define an arm width as the separation across the arm between the mean location of radio masers (near the shock lane) and the mean location of the broad diffuse CO 1-0 gas (near the potential minimum). The galactic spiral 'potential minimum' corresponds to the gas 'surface density maximum' (Gittins & Clarke 2004, their Section 4.2), observable as the broad angular diffuse CO 1-0 gas peaking in intensity when scanning the galactic plane in galactic longitudes across a spiral arm.

**Figure 2** shows the offsets for the maser lane, as measured from the location of the diffuse CO 1-0 gas (CO arm tangents in Table 1, or CO spiral curves in Table 2). Each spiral arm is shown separately, each in a different color. A least-squares fitted line (dashes) is shown. The slope of the dashed line is 25.3 ± 5 pc per kpc. The last point in Figure 2 has been added to reflect a recent paper about masers near the Scutum arm in Galactic Quadrant I located beyond the Galactic Center (Sun et al 2018), as explained below (Section 4).

On the theoretical side, Roberts (1975 – his Fig. 3) predicted each spiral arm to be similar. Thus the streamline starting with a shock would cover a quasi-circular orbit around the Galactic Center, covering a length of roughly $2\pi R_G$, at a mean galactic radius $R_G$. That relative orbital offset in an arm is given as 3.7%, between the shock position and the location of the 'potential minimum' (Roberts, 1975 – his Fig. 2). Thus at $R_G$ = 8 kpc and m=4 spiral arms, that offset would amount to 465 pc (i.e., 0.037 x 2 x 3.1416 x 8000 / 4).

Also on the theoretical side, Gittins & Clarke (2004) looked at the offset from the shock to the 'potential minimum', finding this offset to be a function of the interarm distance (going across the arm), namely 41% (their Fig.13 for 4 arms) or 290 pc for an interarm of 0.7 kpc at $R_G$ = 4 kpc (their Fig. 15), and 36% or 290 pc for an interarm of 0.8 kpc at $R_G$=5 kpc, and 9.6% or 120 pc for an interarm of 1.2 kpc at $R_G$=8 kpc, and 6.4% or 110 pc for an interarm of 1.7 kpc at $R_G$=11 kpc, or 24% or 450 pc for an interarm of 1.9 kpc at $R_G$= 13 kpc.

**Figure 3** shows our results again, but with the addition of the predicted offsets from the density wave theory of Gittins and Clarke (2004). The predicted offset varies with increasing galactic radius, on account of the location of corotation radius (where gas and stars have the same velocity as the density wave spiral pattern).

Thus below a galactic radius of 8 kpc, and above a galactic radius of 12 kpc, the predictions of Gittins & Clarke (2004) are sufficiently close to the observations in Figure 3. In between, the sole observation (in the Perseus arm, near 10 kpc) does not support a corotation radius at 9.6 kpc predicted in the model of Gittins &

Clarke (2004); rather, a corotation beyond 13 kpc is suggested here by the observational data.

The observed radius of corotation in the Milky Way is still being debated: using maser data from the Perseus arm, Vallée (2018a) predicted corotation to be >12 kpc, while the HI data in Foster and Cooper (2010) predicted that corotation is nearer 14 kpc, and the analysis of masers near the Cygnus arm indicated a corotation radius >15 kpc (Vallée 2019).

## 4. New masers in Scutum arm, beyond the Galactic Center in Galactic Quadrant I

Recent publications by Sun et al (2018) and Sanna et al (2017) added a few new masers in that direction beyond the Galactic Center, a direction often referred to as the 'Zona Galactica Incognita' (Fig. 3 in Vallée, 1995; Fig. 2 in Vallée 2002; Fig. 2 in Vallée 2005).

As these masers have not been measured trigonometrically, it was necessary to use a kinematic model to infer their distances to the Sun, and to the Galactic Center.

Basic equations were employed to derive the distance to each maser, in Table 3. From Equation (1) in Roman-Duval et al (2009):

$$R_G = V_{orb} R_{sun} \sin(l) / [v_{rad} + V_{lsr} \sin(l)] \quad (1)$$

where $V_{orb}$ is the orbital circular velocity (taken as 233 km/s for all radii in a flat rotation curve), $V_{lsr}$ is the Local Standard of Rest's orbital velocity (also taken as 233 km/s), $v_{rad}$ is the object's radial velocity (seen at the Sun), $R_G$ is the maser's galactic radius, $R_{sun}$ is the Sun's galactic radius (taken as 8.1 kpc), and $\sin(l)$ is the sinus of the galactic longitude l.

**Figure 4** shows the situation.

After getting $R_G$, to get the solar distance d, one needs the angle m subtended by the object from the Galactic Center, through the equation:

$$\tan(l) = [R_G \cdot \sin(m)] / [R_{sun} + R_G \cdot \cos(m)] \quad (2)$$

and then the distance d follows from the equation:

$$d = [R_G \cdot \sin(m)] / \sin(l) \quad (3)$$

**Table 3** shows the coordinates of these recent masers, employing the kinematical method mentioned above to get the respective distances d and $R_G$.

This path to get d requires some intermediary products. Here we choose to get first $R_G$ from the longitude l and from $v_{rad}$ (equ. 1), and second to get the angle m from $R_G$ and the longitude l (equ.2), and third comes the heliocentric distance d from $R_G$ and from m as well as from the longitude l (equ. 3). None of these equations involves squared quantities.

Elsewhere another path can be employed (Equ. 5-124 in Lang 1980):
$$R_G^2 = R_{sun}^2 + d^2 - 2R_{sun}\, d\, \cos(l)$$
where $R_G$ must be determined first from $v_{rad}$. Both paths are correct.

**Figure 5** shows a map of the Galactic disk toward Galactic Quadrant I, as seen from above the Galaxy. The Scutum arm (in a blue curve), the nearby trigonometric masers (blue filled squares) and the distant kinematic masers (circled blue squares) are shown. Most (5/7) of the masers are inward of the spiral arm, beyond the Galactic Center, closer to the Galactic Center than the diffuse CO 1-0 arm.

The offset from the blue diffuse CO arm and each blue maser beyond the Galactic Center was measured, yielding a mean offset of 535 pc, with a mean galactic radius of 13.5 kpc. This point was added to the data in Figures 2 and 3.

**Figure 6** shows the kinematical map of the Galactic disk toward Galactic Quadrant I, with the radial velocity as a function of galactic longitude. Again, the Scutum arm (in a blue curve), the nearby trigonometric masers (blue squares) and the distant kinematic masers (circled blue squares) are shown. Most (5/7) of the masers are upward from the CO-spiral arm in Galactic Quadrant I.

The offset from the blue CO arm and each maser below the Galactic Center (below the v=0 horizontal line) was measured, yielding a mean offset of 4.5 km/s upward from the spiral arm.

Kinematic distances in the first and second Galactic Quadrants using HI or CO velocities are affected by the non-circular velocity imparted to the HII region by its original formation within the shock, a velocity imprint which it carries downstream as it migrates out of the shock. The reader is warned that the shock could have moved their positions a bit, near the Scutum-Outer (blue) arm in Figure 5 around x=10-12 kpc and in Figure 6 near -80 to -90 km/s.

**5. Discussion on arm width – using a single arm tracer**

Reid et al (2019) fitted log-periodic spiral arms to the location of radio masers in the Milky Way plane, but they had to do something in order to cover apparent gaps in spiral arms without any observed trigonometric (parallax measured) radio masers. So they added some assumptions, namely four 'kinks' in Galactic Quadrant I, one 'kink' in Galactic Quadrant II (their Figure 2), and four 'constrained tangencies' in Galactic Quadrant IV (their Table 2), to slightly bend the pitch angle of a long spiral arm at different galactic radii.

Reid et al (2019 – their Fig. 4) employed another definition of an arm width, namely the width is the intrinsic (Gaussian 1-sigma) scatter in the maser locations near a spiral arm traced only by radio masers. Their definition of such an arm width involves a single tracer (radio masers).

Reid et al (2019) fitted a straight line of this arm width (from one tracer) with increasing galactic radius, and found it to increase along the Galactic radius, at a rate of 36 pc per kpc – we reproduce their fit here in our Figure 3 as 'plus' signs. Again, there is no decrease to zero of their arm width near $R_{Gal}$ = 9.6 kpc (the predicted corotation radius chosen by Gittins & Clarke 2004), thus pushing the corotation radius beyond 13 kpc.

The use of the 'scatter in maser locations', to represent an arm width, is problematic. It ignores those radio masers located in the interarm regions, thus accidentally bending the long spiral arm and its pitch angle (Vallée 2017b) and artificially enlarging the scatter (thus the arm width).

False growth in the scatter with distance? The scatter itself used by Reid et al 2019 to measure the arm width (as the scatter in data points) should grow with the distance from the Sun, as there is a systematic bias introduced by the inversion of the parallax value (the scatter increasing with distance due to an uncorrected inversion – see Bailer-Jones 2015). The scatter in maser distances should grow with increasing distance from the Sun, due to the instrumental errors (hence an artificial bias, not a physically real growth); ditto for a growth in scatter for an increasing distance from the Galactic Center (artificial, not physically real). Hence the apparent agreement of their maser width in Figure 3 with our two-tracer offset may be mostly fortuitous.

Also, some masers were observed to change their size over time, such as Sharpless 269, leading to different distance estimates separated by 1 kpc – an improper size assumption would increase the scatter value with increase distance from the Sun. The various different parallax distances to Sharpless 269 (Honma et al 2007 found 5.3 ±0.2 kpc, Asaki et al 2014 found 4.0 ±0.2 kpc, and Quiroga-Nunez et al 2019 found 4.2 ±0.2 kpc) which seems to be due in part to the assumed linear size (zero for a compact core, versus a complex morphology – Quiroga-Nunez et al 2019). In addition, a maser's short periodic motion on the sky (< 1 year) could confuse a parallax estimate.

Hence it is possible that the scatter in maser location, to represent an arm width with galactic radius, may be grossly inflated with distance, by some interarm masers, by an uncorrected inversion, by an improper size assumption, and by instrumental errors with distance; all these would diminish any agreement with our own 2-tracer arm width in our Figure 3.

Finally, the use of a single tracer to define an arm width ignores the arm's other tracers (dust, CO, HII regions, old stars, etc) and the physical offsets of other tracers from the masers' locations.   Figure 7 shows some observed tracers in Galactic Quadrant IV (CO in blue, MIR dust in red, radio HII complexes in orange), when compared to the predictions of  the trigonometric maser-based model of Reid et al (2019). The aging gradient is well seen (from the right to the left) in each spiral arm. The Reid et al model in Galactic Quadrant IV (vertical dashed lines) appears to be ad hoc, as there is no parallactic maser observed in that area.

There are some 'kinematical' masers (*not measured* with parallaxes), peaking near $l=312^o$ in the Scutum arm, and near $l=330^o$ in the Norma arm (see Table 7 in Vallée 2016a), and both observed longitudes (in green) are closer to the dust lane in each arm than to the prediction in the model of Reid et al (2019).

## 6. Multi-tracer approach to defining the spiral arm width in nearby spiral galaxies

In nearby spiral galaxies, the two-tracer approach has already been employed – for a recent review, see Vallée (2020).  In nearby spiral galaxies, the width of a spiral arm is usually measured between two tracers, namely one tracer in the starforming regions (choice of  H-alpha, 3.6-micron emission, Blue band, etc) and one other tracer way beyond the dust lane (choice of old stars, old star clusters, HI atom, CO gas, etc).

Thus, in 24 nearby spiral galaxies, recent statistics indicated the separation of the dust lane (or a nearby tracer – see above) from the older stars (or a nearby tracer such as HI – see above) to have a median offset value near 326 ± 50 pc (Vallée 2020).

Interferometric maps of nearby spiral galaxies may not show the large-scale diffuse cold molecular gas, because of missing small interferometric baselines (see Section 2.1). Hence such interferometric maps will only detect the clumpy CO molecular gas, located nearer the young optical HII regions and the dust lanes.

The presence of a measured arm width, *beyond* 6 *sigma* (326/50) in nearby spiral galaxies, as observed using two independent tracers (one in the starforming regions, one from the region of older stars), attest to the physical reality of this tracer offset. It also bolsters theories of arm formation that predict such an offset (density-waves), and questions the theories that predict zero offset (tidal theories, dynamic spiral – Dobbs & Baba 2014, etc).

Similarly, in the Milky Way, the large compilation of all observed tangents to each arm, in many different tracers (Vallée 2016a – tables 1 to 10) has shown statistically the presence of a measured mean arm width in Section 4.1 in Vallee (2017d): combining all the inner galactic spiral arms, the angular separation between the dust tracer and the large-scale diffuse CO 1-0 gas tracer at a typical solar distance of 5.6 kpc was found to be 3.2° with an r.m.s. of 0.68° and a s.d.m. of 0.3°, while the linear separation was found to be 315 pc with an r.m.s. of 64 pc and a s.d.m. of 26 pc, thus *beyond* a *sigma* (signal/sdm) of 12.

### 7. Conclusion

In each spiral arm, we have assembled the observed separation between the mean location of the radio masers and the mean location of the diffuse CO 1-0 gas intensity peak (Table 1 and Table 2; Figure 1). We then investigate this separation (masers versus diffuse CO peak) with galactic radius (Figure 2).

Here we investigate a multi-tracer approach to defining the spiral arm width, in the Milky Way. There is already a large catalogue of all observed tangents to each spiral arm, as done before with many different observational tracers (Vallée 2016a – tables 1 to 10).

Our main conclusions are as follows:

- In the Milky Way disk, the separation of radio masers (shocked gas near the inner edge of a spiral arm) and the aging stars and diffuse CO 1-0 gas ('potential minimum' near the arm middle) is 250 ± 50 pc – see Fig. 1 and Tables 1 and 2.

- This maser-to-diffuse CO 1-0 gas separation value increases with the Galactic radii of the radio masers, typically being around 25 ± 5 pc per kpc – see Fig. 2.

- There is no sign of a decrease of this separation (masers versus diffuse CO gas peak), which would occur as one approaches corotation of gas and spiral pattern, suggesting that corotation is beyond a galactic radius of 12 kpc – see Fig. 3.

- The observed values of separation and its increase with galactic radius are consonant with their predictions within the density wave theory, before corotation (Section 1).

Here we have employed published diffuse CO 1-0 gas data and arm model and older data (Tables 1 and 2) in addition to newer data from radio masers (Table 3, Fig. 4, Fig. 5 and 6), in order to investigate their positions within their own spiral arms.

Figure 2 shows the arm width so defined by the separation between two tracer offsets (masers and diffuse CO). This separation found here near 250 ± 50 pc between the maser locations and the diffuse CO gas peak location is somewhat smaller than the previously determined 315 ± 30 pc separation between the dust lane and the diffuse CO gas peak location (Vallée 2016c; Vallée 2016a).

Arm width. Any combination of two tracers can be employed, provided one tracer is in the starforming region (near the 'shock') and the other tracer is in the aged stars regime (near the 'potential minimum'). An arm width should encompass both the region of dust and starforming regions, as well as the region of aged stars and diffuse CO gas; thus it should not be measured by a single arm tracer (e.g., just the radio masers) and this width should not be impacted by some interarm matter (variously called armlets, blobs, branches, bridges, feathers, fingers, kinks, segments, spurs, sub-arms, swaths, etc ) – see Vallée (2018b).

In Section 5 and Figure 3, we compare our two-tracer arm results with a one-tracer (maser-based) arm model of the spiral arms in the Milky Way (Reid et al 2019) and point out some concerns when using the 'scatter in maser locations' to represent the arm 'width', in a single-tracer arm model. Also, Figure 7 shows concerning discrepancies between the model prediction of the 'location' of masers of Reid et al (2019) and some observed arm tracers. The use of a single tracer, to get both the arm location and the arm width, is most problematic.

**Acknowledgements**
The figure production made use of the PGPLOT software at NRC Canada in Victoria. I thank an anonymous referee for precise, useful, careful, and historical suggestions.

Table 1.  Arm tangents in Galactic longitudes, for different arm tracers (broad diffuse CO gas and radio maser lane), as observed tangentially from the Sun's position

| Spiral arm | Gal. longit.[a] of tangent to diffuse $^{12}$CO 1-0 arm tracer (in a 8.8' HPBW) (deg) | Gal. longit.[a] of tangent to methanol arm tracer (deg) | CO-to-maser offset, at solar distance $r_o$ (deg; pc, kpc) | CO-to-maser direction[c] toward or away from Gal. Center | Gal. radius of arm $r_G$, at offset (kpc) |
|---|---|---|---|---|---|
| Carina-Sagittarius | 281.3 | 284.5 | -3.2°; 251 pc at 4.5 kpc | toward GC | 8.0 |
| Crux-Scutum | 309.5 | 312.2 | -2.7°; 259 pc at 5.5 kpc | toward GC | 6.0 |
| Norma | 328.4 | 330.4 | -2.0°; 202 pc at 5.8 kpc | toward GC | 4.2 |
| Perseus start | 336.8 | 337.8 | -1.0°; 119 pc at 6.8 kpc | toward GC | 3.2 |
| Sagittarius start[b] | 343.0 | 344.0 | -1.0°; 131 pc at 7.5 kpc | toward GC | 2.5 |
| Norma start[b] | 019.0 | 016.0 | +3.0°; 314 pc at 6.0 kpc | toward GC | 2.5 |
| Scutum | 032.9 | 028.5 | +4.4°; 414 pc at 5.4 kpc | toward GC | 4.0 |
| Sagittarius | 050.5 | 050.0 | +0.5°; 042 pc at 4.8 kpc | toward GC | 6.2 |

Notes. (a): Mean galactic longitude data come from Table 3 in Vallée 2016a; individual data stem from Tables 5 and 7 in that same reference.

(b) For Sagittarius start, and Norma start, data from Table 4 in Vallée 2016b.

(c) Offset direction (from diffuse CO tracer to radio maser lane) is always pointing toward the direction of the Galactic Center.

Table 2. Arm tracer distances located close to the Galactic Meridian, for different arm tracers (diffuse CO gas and radio maser lane)

| Spiral arm | Distance (a) from Sun to diffuse $^{12}$CO 1-0 model (kpc) | Distance (a) from Sun to methanol maser lane (kpc) | CO-to-maser offset (b) | CO-to-maser direction, toward or away from Gal. Center | Gal. radius of arm (kpc) |
|---|---|---|---|---|---|
| Perseus l=180º | 2.6 | 2.3 | +310 pc | toward GC | 10.6 |
| Sagittarius l=0º | 0.7 | 1.1 | -380 pc | toward GC | 7.3 |
| Scutum l=0º | 3.0 | 3.2 | -190 pc | toward GC | 5.0 |
| Norma l=0º | 4.5 | 4.6 | -120 pc | toward GC | 3.5 |

Notes. (a): Mean distance data come from Table 7 in Vallée (2019); individual data stem from Figure 1 in that same reference.

(b) Offset direction (from diffuse CO tracer to radio maser lane) is always pointing toward the direction of the Galactic Center.

Table 3. Masers found near the Scutum arm beyond the Galactic Center, in Galactic Quadrant I

| Name | Gal. Long. (o) | Gal. Lat. (o) | Cloud V$_{lsr}$ (km/s) | Reference | Solar distance (kpc) | Gal. radial dist. (kpc) | Note on dist. |
|---|---|---|---|---|---|---|---|
| G007.47+0.05 | 007.47 | +0.05 | -16.0 | Sanna et al 2017 | 20.4 | 12.5 | parallax dist. |
| G28.32+1.24 | 28.32 | +1.24 | -44.8 | Sun et al 2018 | 20.1 | 13.6 | kin. model |
| G34.84-0.95 | 34.84 | -0.95 | -45.1 | Sun et al 2018 | 17.9 | 12.2 | kin. model |
| G39.18-1.43 | 39.18 | -1.43 | -55.6 | Sun et al 2018 | 18.1 | 13.0 | kin, model |
| G40.29+1.15 | 40.29 | +1.15 | -50.6 | Sun et al 2018 | 17.1 | 12.1 | kin. model |
| G40.96+2.48 | 40.96 | +2.48 | -59.3 | Sun et al 2018 | 18.1 | 13.2 | kin. model |
| G55.11+2.42 | 55.11 | +2.42 | -68.3 | Sun et al 2018 | 15.3 | 12.5 | kin. model |

Note: the kinematical model employed is that of Vallée (2017a, 2017b), with 8.1 kpc as the distance of the Sun to Galactic Center, and 233 km/s as the orbital velocity of the Local Standard of Rest.

**Figure Captions**

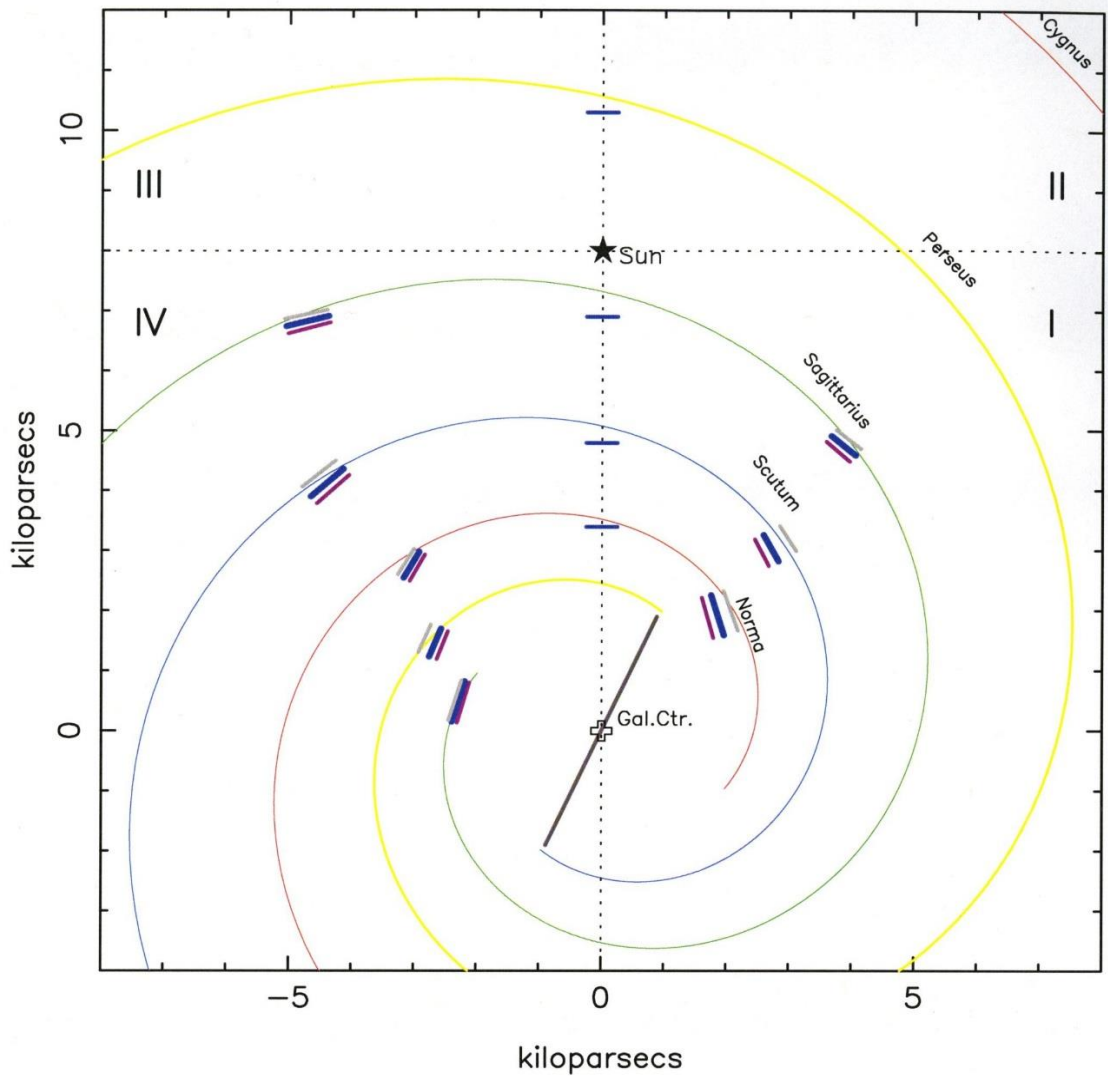

**Figure 1.** The galactic disk seen face-on. The Sun's location (filled star) and the location of the Galactic Center (open plus sign) are shown. Galactic Quadrants I to IV are indicated. Shown are the long arms: Norma-Cygnus arm (red), Perseus arm (yellow), Sagittarius-Carina arm (green), and Scutum-Crux-Centaurus arm (blue). One can see the separation between the radio masers (blue bars), the diffuse CO 1-0 gas intensity tangents (gray bars), and the Near and Mid Infrared measurements of dust (Table 2 and Fig. 1 in Vallée 2016a; Table 2 and Fig. 4 in Vallée 2017d – red bars), as well as the 4-arm global spiral model previously fitted to the broad diffuse CO 1-0 arm tangents. Both masers (blue bars) and dust (red bars) are always located inward from the diffuse CO peaks (gray bars), towards the Galactic Center.

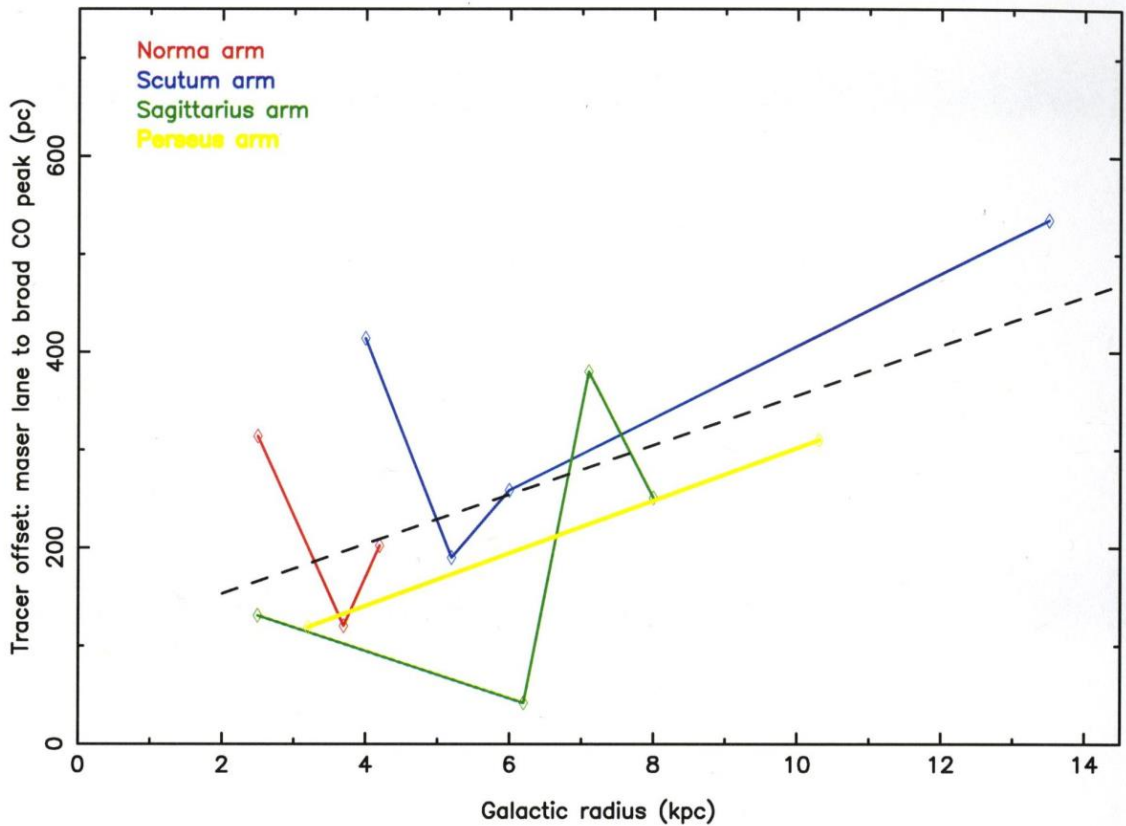

Figure 2. The tracer offset (maser lane versus diffuse CO gas peak) in a spiral arm (vertical), as a function of the arm's distance from the Galactic Center (horizontal). Different spiral arms are shown in different colors: Perseus (yellow), Norma-Cygnus (red), Sagittarius-Carina (green), Scutum-Outer (blue). A rough sketch (dashed line) is shown of the offset increase with galactic radius, from a least-squares-fit to the observational data. Data from Tables 1 (8 points) and 2 (4 points), and a last point (top right) from Table 3 and Section 4 (13.5 kpc, 535 pc).

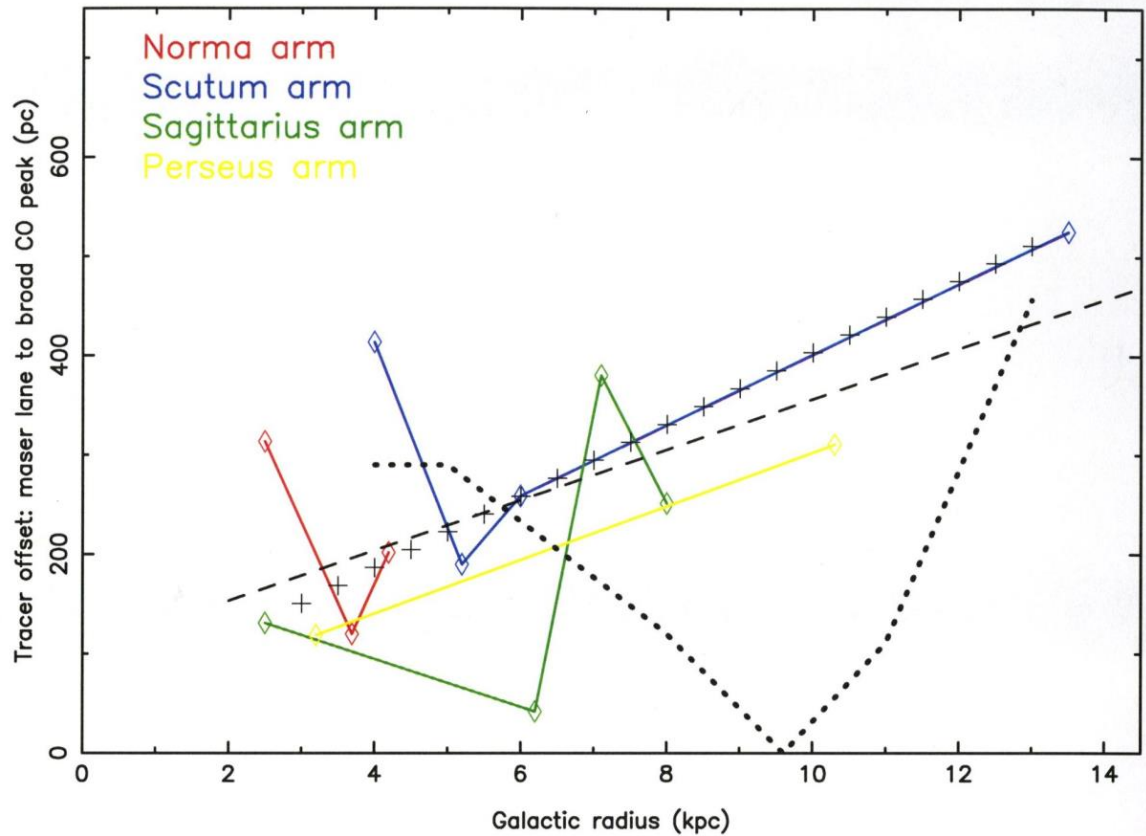

Fig. 3 – Same 13 observational data as in the previous figure (four colored arms, dashed least-squares-fitted line), with the addition of the predicted offsets (shock to spiral 'potential minimum') from the density wave model of Gittins & Clarke (2004, their figures 13 and 14) – see the dotted descending and ascending curve. Their model assumes a corotation (of gas and stars with the spiral pattern) at a galactic radius near 9.6 kpc, and hence their offset must go to 0 there. The arm width of Reid et al (2019) is shown (+ signs) - see text in Section 5.

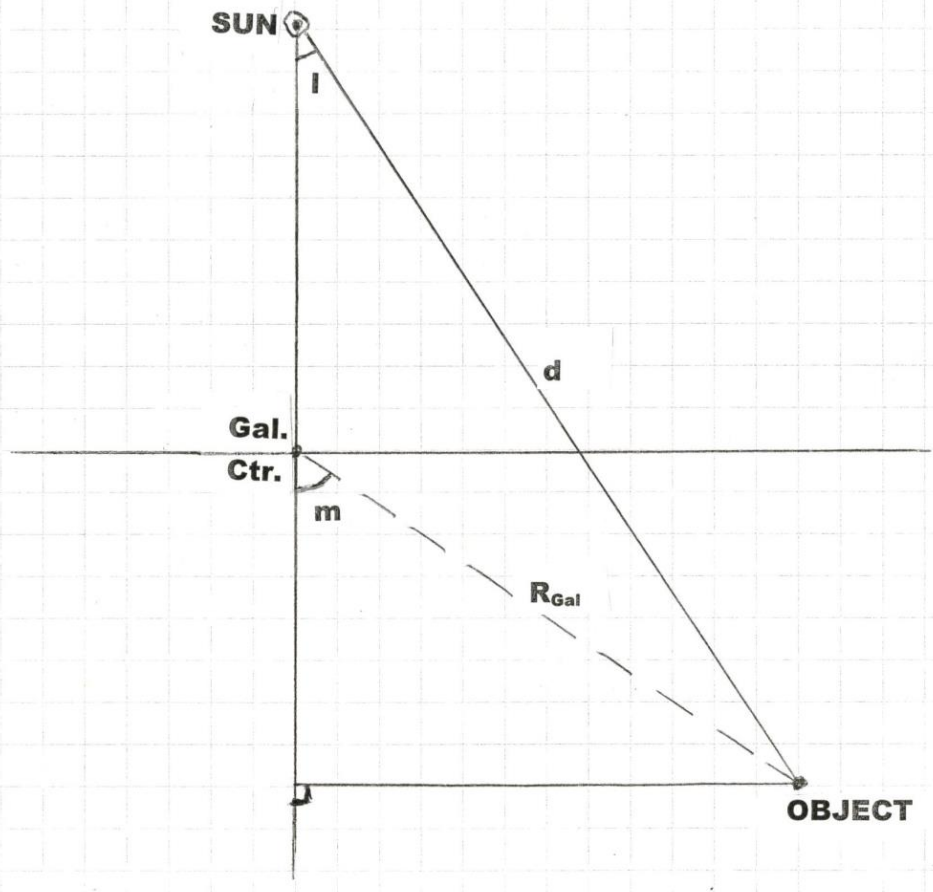

Figure 4. A rough sketch of the geometry for the equations 1 to 3. The Sun (circled dot), the Galactic Center (Gal. Ctr.) and the object (OBJECT), along with the galactlc longitude (l), and the view angle from the Galactic Center (m). The distance from the Galactic Center is labeled $R_{GAL}$, and the distance from the Sun is labeled d.

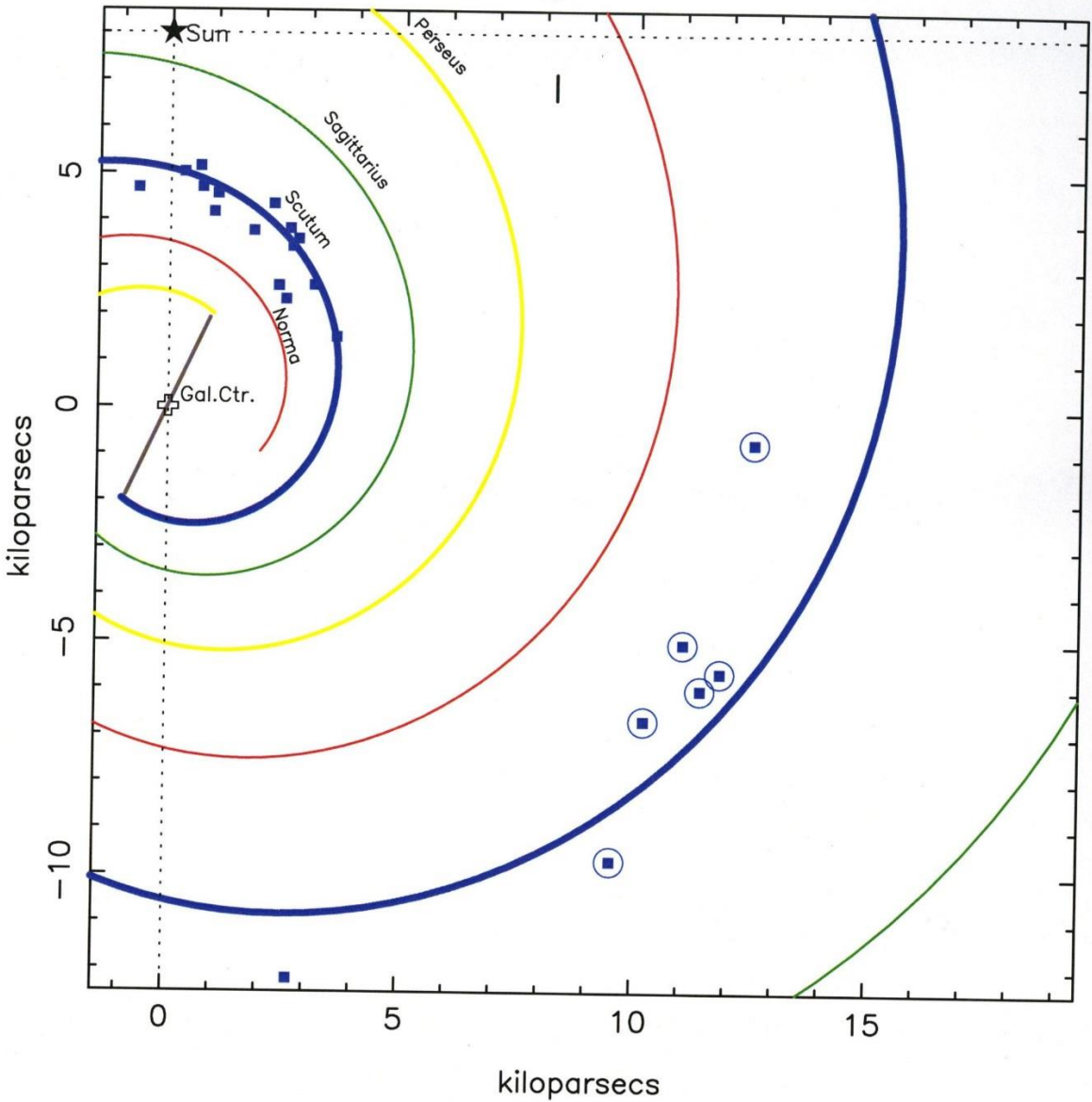

**Figure 5.** Map of Galactic Quadrant I of the Galactic disk, as seen from above. The Scutum arm is shown in blue. Different spiral arms are shown in different colors: Perseus (yellow), Norma-Cygnus (red), Sagittarius-Carina (green), Scutum-Outer (blue). The masers are shown in squares. Data from Table 3 here.

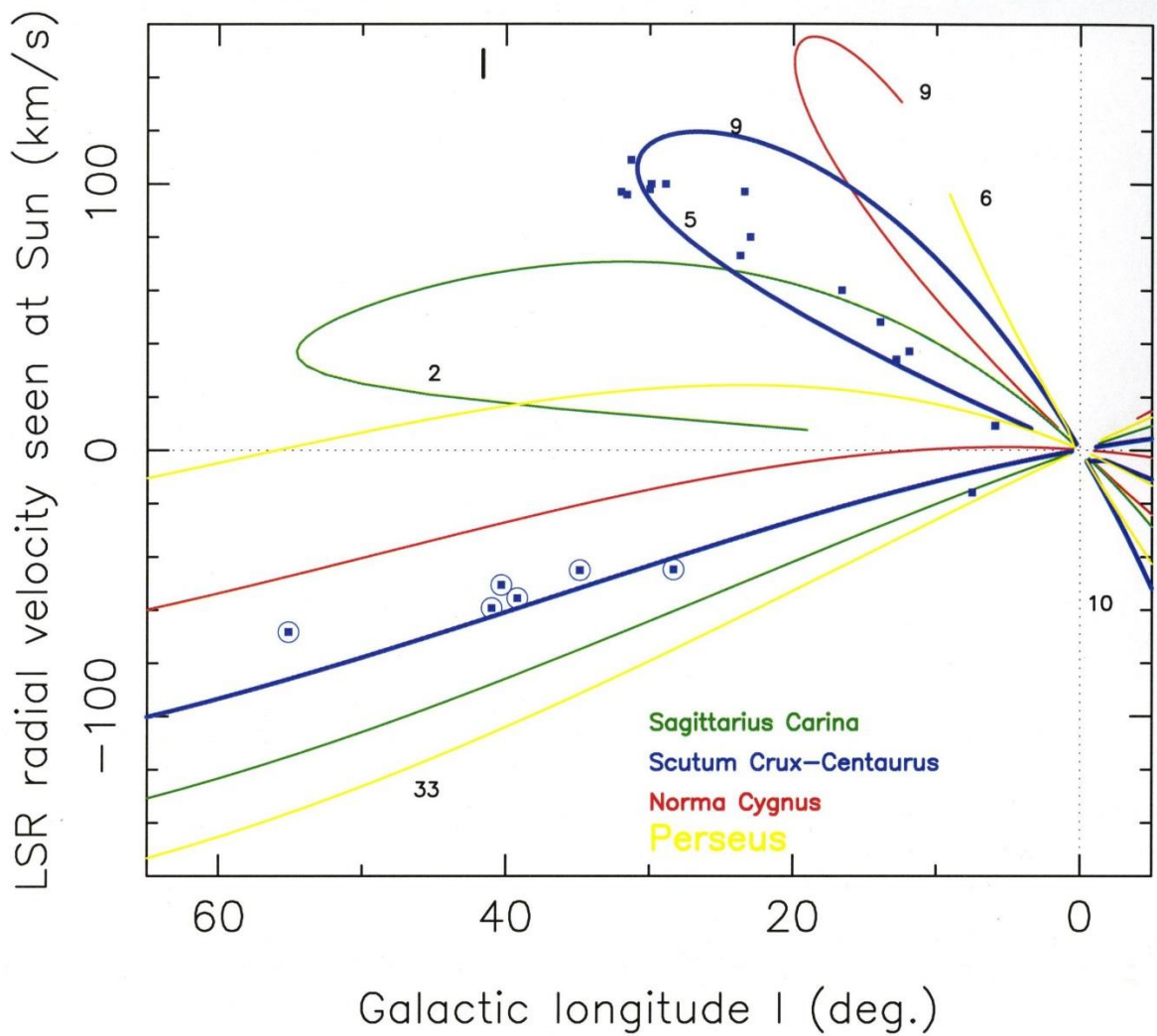

**Figure 6.** Kinematical map of Galactic Quadrant I of the Galactic disk, with galactic longitude (horizontal) and radial velocity (vertical), as seen from the Local Standard of Rest (LSR). The Scutum arm is shown in blue. The LSR's orbital circular velocity around the Galactic Center is 233 km/s (Vallée 2017c). Different spiral arms are shown in different colors: Perseus (yellow), Norma-Cygnus (red), Sagittarius-Carina (green), Scutum-Outer (blue). The masers are shown in squares. Data from Table 3 here.

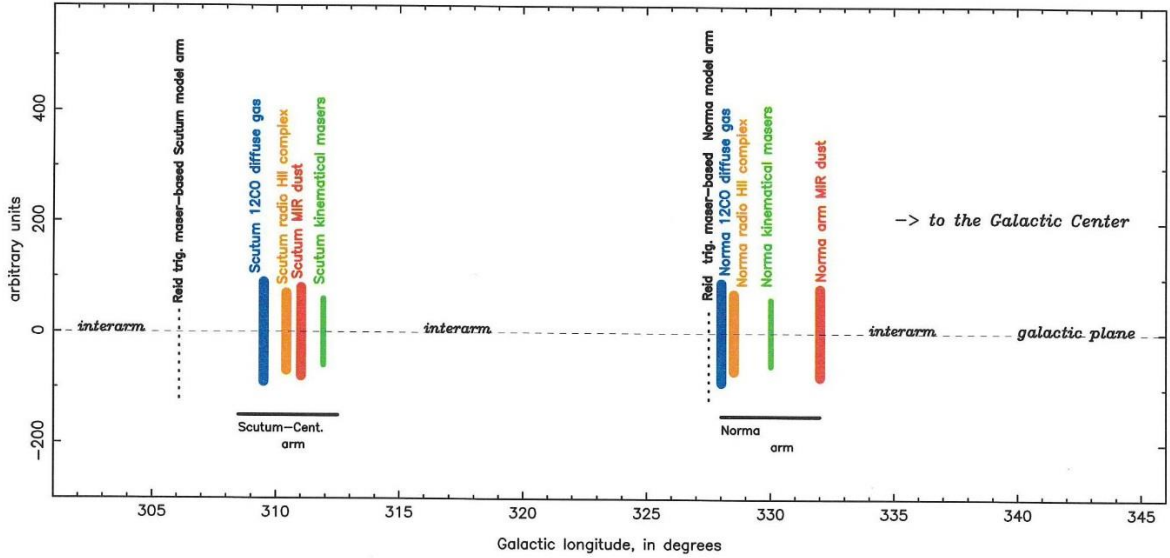

Figure 7. Radio-to-infrared observational view of the arm tangents in galactic longitude versus two spiral arms in Galactic Quadrant IV.  Masers should be near the dust lane region and the very young stars (closer to the direction toward the Galactic Center),     while old stars should be near the broad diffuse CO 1-0 peak. Errors in galactic longitude for typical arm tracers are around 0.9°, while an arm width in this quadrant is around 4°.

*At left*, one sees the Scutum-Centaurus arm tangents: the observations of the diffuse CO 1-0 gas is peaking at l=309.5° (Table 5 in Vallée 2016a), observed radio HII regions at l=310.4° (Table 6 in Vallée 2016a), while the observations of the 60µm and 240µm dust is peaking at l=311° (Table 3 in Vallée 2016a). The maser-based model of Reid et al (2019) appears at l= 306.1° (their Table 2).  The offset, between the maser-based model of Reid et al (2019) and the MIR dust peak tracers, is 3.4° (or 356 pc at a solar distance of 6 kpc) and way away from the inner arm edge.

*In the middle*, one sees the Norma arm tangents:   the observations of the diffuse CO 1-0 is peaking at l=328° (Table 5 in Vallée 2016a), observed radio HII regions at l=328.1° (Table 6 in Vallée 2016a),  while the observations of the 240µm and 2.4µm  dust is peaking at l=332° (Table 3 in Vallée 2016a). The trigonometric maser-based model of Reid et al (2019) appears at l= 327.5° (their Table 2). The offset, between the maser-based model of Reid et al (2019) and the MIR dust peak tracers, is  4.5° (or 550 pc at a solar distance of 7 kpc) and way away from the inner arm edge.